\documentclass[preprint]{aastex}
\usepackage{graphicx}
\usepackage{natbib}
\newcommand{\finv}{{F_{\rm inv}}}
\newcommand{\tn}{{t_{\rm nuclear}}}
\newcommand{\thn}{{\thinspace}}
\newcommand{\el}[2]{\ensuremath{^{#1}\mathrm{#2}}}
\newcommand{\Mo}{\rm{M}_\odot}

\def\mnras{MNRAS}

\def\apj{ApJ}
\def\aap{A\&A}
\def\aj{AJ}
\def\pasp{PASP}
\def\apjl{ApJ}
\def\apjs{ApJS}

\shorttitle{Thermohaline Mixing in M3}
\shortauthors{Angelou et al.}

\title{Thermohaline Mixing and its Role in the
  Evolution of Carbon and Nitrogen Abundances in Globular Cluster Red Giants: The Test Case of Messier 3}
\author{George C. Angelou\altaffilmark{1}, Ross P. Church\altaffilmark{2,1}, Richard J. Stancliffe\altaffilmark{1}, John C. Lattanzio\altaffilmark{1}, Graeme H. Smith\altaffilmark{3}}
\altaffiltext{1}{Centre for Stellar and Planetary Astrophysics,  School of Mathematical Sciences, Monash University,  Melbourne,  VIC 3800,  Australia.}
\altaffiltext{2}{Lund Observatory, Box 43, SE-221 00 Lund, Sweden.}
\altaffiltext{3}{University of California Observatories/Lick Observatory, Department of Astronomy and Astrophysics, UC Santa Cruz,1156 High St., Santa Cruz, CA 95064, USA.}

\email{George.Angelou@monash.edu}
\begin{document}

\begin{abstract}
We review the observational evidence for extra mixing in stars on the red 
giant branch (RGB) and discuss why thermohaline mixing is a strong
candidate mechanism. 
We recall the simple phenomenological description of thermohaline mixing, and 
aspects of mixing in stars in general. We use observations of M3 to 
constrain the form of the thermohaline diffusion coefficient and 
any associated free parameters. This is done by matching [C/Fe] 
and [N/Fe] along the RGB of M3. After taking into account a presumed 
initial primordial bimodality of [C/Fe] in the CN-weak and CN-strong 
stars our thermohaline mixing models can explain the full spread 
of [C/Fe]. Thermohaline mixing can produce a significant change in 
[N/Fe] as a function of absolute magnitude on the RGB for initially 
CN-weak stars, but not for initially CN-strong stars, which have so 
much nitrogen to begin with that any extra mixing does not significantly 
affect the surface nitrogen composition.
\end{abstract}


\section{Introduction}

Standard stellar evolution theory predicts that only one mixing event will
change the surface composition of a low-mass star as it ascends the red
giant branch. That event is the so-called first dredge-up (FDU, see \citealt{1967ApJ...147..624I}) associated with
the inwards migration of the base of the convective envelope into regions where
hydrogen burning via the CNO-bicycle has occurred. Relatively modest
changes in surface C and N abundance are predicted, and once the convective envelope
recedes outwards these changes are brought to a halt. However, observations
of low-mass red giants ($M < 2.5 M_\odot$, see for example \citealt{1998A&A...336..915C}) for essentially all compositions show
trends among light element abundances that cannot be accounted for by the
FDU. Some form of non-convective mixing seems to occur whereby greater
amounts of the products of partial hydrogen burning are cycled into the convective
envelope over a much longer timescale, and during more advanced phases of
RGB evolution, than can be explained by the FDU. The observational results
summarised
below imply that current canonical models do not include essential physics
of this so-called ``extra mixing''. Regardless of what the physical mechanism
is, extra mixing is required to conform to the following
observational criteria:

\begin{enumerate}

 \item It commences after the hydrogen burning shell has erased a
   composition
 discontinuity in the radiative zone that marked the innermost limit of the
 convective envelope during the FDU event, and may continue to at
 least the tip of the RGB \citep{1991ApJ...371..578G, 1998A&A...332..204C, 2000A&A...354..169G, 2003PASP..115.1211S, 2003ApJ...585L..45S, 2004MmSAI..75..347W,  2008AJ....136.2522M}. The onset of the extra
 mixing is thus thought to coincide with a local maximum (the so-called
 ``bump'')  observed in the RGB luminosity function of globular clusters.

\item It must occur over a range of masses and metallicities
\citep[and references therein]{2009A&A...502..267S},
being active in giants of all metallicities from
solar to at least [Fe/H] $\sim -2.5$ \citep{2000A&A...354..169G} and masses less than
$\sim 2.5 M_\odot$ \citep{1977ApJ...217..508L}, although not necessarily with equal
efficiency throughout these mass and metallicity ranges.

 \item It must deplete \el{7}{Li} \citep{1998A&A...332..204C, 2009A&A...502..267S,2009A&A...503..545L}.

 \item It must decrease the \el{12}{C}/\el{13}{C} ratio \citep{1994A&A...282..811C, 1996ASPC...98..213C}, since values
lower than predicted by the FDU are found among Population I field giants
\citep{1976ApJ...210..694T, 1981ApJ...248..228L, 1998A&A...332..204C}, open
cluster giants \citep{1989ApJ...347..835G, 1991ApJ...371..578G,2009A&A...502..267S, 2010MNRAS.407.1866M}, globular cluster giants \citep{2003ApJ...585L..45S, 2007A&A...461L..13R} and halo field giants \citep{1986ApJ...311..826S,2000A&A...354..169G}.

\item It must decrease the total carbon abundance since systematic
  decreases  with advancing luminosity on the upper half of the red giant branch are
seen both among globular clusters and halo field giants \citep{1981ApJS...47....1S,1989asgc.conf...71S,1982ApJS...49..207C,
1983ApJ...266..144T,1986PASP...98..473L,2000A&A...354..169G,2001PASP..113..326B,2003PASP..115.1211S,2008AJ....136.2522M,2010AJ....140.1119S}.
In Population II giants the behaviour of the carbon abundance can serve
as an even more potent probe of the extent of extra mixing than the \el{12}{C}/\el{13}{C}
isotope  ratio, because the latter can attain near-equilibrium values for only
moderate amounts of mixing that would otherwise cause only small ($\sim$0.1 dex)
changes  in [C/H] \citep{1986ApJ...311..826S}.

\item As a consequence of the previous point it must increase the nitrogen abundance. The results of CN cycling are observed on the upper half of the red giant branch. Halo field stars on the upper RGB were found by \citet{2000A&A...354..169G} to show an excess of nitrogen compared to those on the lower RGB.

\end{enumerate}

It is expected that the mechanism(s) will also destroy \el{3}{He} inside the star \citep{1986ApJ...302...35D, 1995PhRvL..75.3977H, 1996ApJ...465..887D, 1999ApJ...510..217S, 2007A&A...467L..15C, 2007A&A...476L..29C}. As we cannot  observe \el{3}{He} in stellar atmospheres directly this is not an observational constraint, but it is a significant requirement from the study of chemical yields and galactic evolution. The importance of \el{3}{He} is discussed in Section 3.

In the last four decades many candidate extra-mixing mechanisms have been suggested.    
These include:  rotational mixing \citep{1979ApJ...229..624S, 2005ApJ...631..540C, 2006A&A...453..261P}, 
magnetic fields \citep{2009PASA...26..161P, 2008ApJ...684L..29N, 2007ApJ...671..802B}, and 
internal gravity waves \citep{2000MNRAS.316..395D}. Individually, none of these have been proven to be satisfactory.  
Due to its promising ability to account for the above requirements, and the necessity of
its occurrence in low mass giants just after the FDU, in this 
study we focus on \el{3}{He}-driven ``thermohaline mixing'' (\citealt[EDL06 hereafter]{2006Sci...314.1580E}, \citealt[CZ07a hereafter]{2007A&A...467L..15C}, \citealt[EDL08 hereafter]{2008ApJ...677..581E}).\footnote{One must concede that multiple processes and indeed interactions between them can affect the transport. Models have been made that include multiple processes \citep{2010A&A...521A...9C, 2010A&A...522A..10C} by simply adding the diffusion coefficients for each process. This does not allow for the interaction between the processes as discussed by \citet{2008ApJ...684..626D}.} The name comes from
a phenomenon seen in oceans, and is taken from the two major determinants of the
density of sea water - its heat content (``thermo'') and its salinity 
(the salt or ``haline'' content). It is common to find warm salty water 
overlying cool fresh  water. Although the higher salinity of the warm water makes it denser, the higher
heat content acts to stabilise the stratification. The subsequent evolution of the system
is determined by the competition between two diffusion processes and their associated 
timescales - the time for the (stabilizing) heat to diffuse away compared to the
time for the (destabilizing) salt to do the same. Hence the process is often called ``doubly-diffusive'',
and it has been studied in the oceanographic context for many years (see recent reviews, 
theoretical modelling, observations and laboratory experiments in
Progress of Oceanography Volume 56, 2003; e.g. \citealt{ruddick}). Within oceans
it is now well known that the rapid diffusion of heat from the warm salty layer produces
an over-dense layer that begins to sink into the cooler fresh water below. The
temperature stays roughly the same as the surrounds, and ``salt fingers'' form which
extend downward delivering the saltier water to deeper regions. Reciprocal fresh-water fingers move upward and replace
the salty water with fresher water. Laboratory experiments have also played a role in helping to characterise the instability. Work by Stommel and Faller published in \citet{stern} as well as more recently \citet[and references there in]{Krishnamurti} have helped elucidate the instability.

A similar process can occur in stars. Here it is not salt but the mean molecular weight
that is the ``destabilising agent'', in the words of \citet{2010arXiv1006.5481D}. Usually, the 
molecular weight increases as we move toward the centre of the star, as a result of 
nuclear burning and fusion reactions. If it were to decrease, then the plasma would be
buoyantly unstable, just as is the case in convection. However, just as in the oceanic case,
we must include the rapid thermal diffusion which can act to stabilise the
motion. In this paper we
consider the situation where some local event causes a 
decrease in the molecular weight in an otherwise stable region within a star.
We investigate the effects of the resultant ``thermohaline mixing'' or doubly 
diffusive process that is initiated by a molecular weight ($\mu$) inversion 
\citep*{1972ApJ...172..165U,1980A&A....91..175K}\footnote{This has been referred
to as ``$\delta \mu$ mixing'' by EDL06 and EDL08, to distinguish it from 
a separate occurrence of ``thermohaline mixing'' in stars. In 
that case we may have mass transfer in a binary system, where material of a higher 
mean molecular weight is accreted on an envelope of lower molecular weight. This situation is
unstable and some thermohaline circulation will take place to redistribute the composition of
the star to result in a stable stratification (exchanging energy with the thermal content 
of the material in doing so). This case is not relevant to the discussions
in this paper, and more information may be found in \citet{2004MNRAS.355.1182C} and
\citet{2007A&A...464L..57S}.}.

Recently thermohaline mixing has featured prominently in the literature. EDL06, 
CZ07a, EDL08, \citet{2010A&A...521A...9C} and \citet{2010A&A...522A..10C} have discussed in detail the 
important consequences of its inclusion during the RGB. 
The dichotomy in RGB carbon abundance between metal poor stars and carbon-enhanced 
metal-poor stars has been explained by \citet{2009MNRAS.396.2313S} using thermohaline mixing. 
\citet{2010A&A...521A...9C}, \citet{2010MNRAS.403..505S} and \citet{2010A&A...522A..10C} have shown that its 
operation beyond the giant branch
may affect the subsequent asymptotic giant branch (AGB) evolution. This mechanism may 
be a crucial part of stellar physics that has been missing from the models. 
We thus believe it pertinent to investigate the effects of the mechanism in some detail.

In this paper we will take the approach of trying to model the mixing
in spherically symmetric stellar models and investigate its effect on observable surface abundances. In
this respect our approach follows that of CZ07a and \citet{2010A&A...522A..10C}. 
We will examine the change of the abundances
of carbon and nitrogen on the red giant branch of globular clusters, focusing on the
case of M3.

Currently a range of approaches is taken to include thermohaline mixing 
in evolution codes, especially if the \el{12}{C}/\el{13}{C} ratio is the 
constraint used to determine
the extent of extra mixing. The  \el{12}{C}/\el{13}{C} ratio was one of the first 
indicators that extra mixing must operate on the RGB.  It is classically used to 
probe the results of FDU \citep{1975MNRAS.170P...7D,1976ApJ...210..694T, 1994A&A...282..811C}. It naturally 
follows that the \el{12}{C}/\el{13}{C} ratio could be used to trace the extent of extra 
mixing and constrain any mechanism. The change in \el{12}{C}/\el{13}{C} ratio 
following FDU will depend on the efficiency of mixing and allow us to explore the mixing 
velocity via a diffusion approximation. EDL08 estimated 
the mixing speed with their formula for the diffusion coefficient and found that a range 
of three orders of magnitude  in their free parameter can lead to the observed levels 
of \el{12}{C}/\el{13}{C} and \el{3}{He} depletion. \citet{1972ApJ...172..165U} and 
\cite*{1980A&A....91..175K} both use essentially the same formula (UKRT formula hereafter) 
for the diffusion coefficient but their choice of the free parameter varies by  
two orders of magnitude. In an attempt to constrain the parameter space we address the 
following questions:
\begin{enumerate}
 \item The \el{12}{C}/\el{13}{C} ratio is generally used as a tracer to probe the
extent of mixing. This ratio saturates near the CN equilibrium value rather quickly in 
low metallicity stars, and hence is of limited utility. Is there a better way to 
constrain the mixing? 
 \item Which formalism should be used? Here we will limit our investigation to
the EDL08 and UKRT prescriptions for the diffusion coefficient. 
 \item Once the preferred formalism is identified, what diffusion
   co-efficient (or mixing velocity) is needed
to match observations? What values do we use for any free parameters?
 
\end{enumerate}

As has been the practice for many years, we turn to globular clusters 
to test our understanding of stellar theory. 
\citet{2002PASP..114.1097S}, \citet{2003PASP..115.1211S} and \citet{2008AJ....136.2522M} have compiled 
observations of carbon and nitrogen along the giant branch of M3. This has provided 
us with a valuable alternative to the \el{12}{C}/\el{13}{C} probe. By matching 
our models to the carbon depletion (as a function of absolute magnitude) observed 
in this cluster we can attempt to constrain both the form of the thermohaline diffusion 
coefficient and  the values of any parameters contained therein. As carbon and 
nitrogen are intrinsically linked in the CN burning cycle we include observations 
of nitrogen as an additional tracer. Furthermore, we identify when extra mixing begins 
in the models and compare this to the observed luminosity function bump (LF bump) 
in the cluster.

\section{Thermohaline Mixing}
The usual condition for convective instability is simply that a blob moved from
its equilibrium position will be buoyantly unstable and continue to move away from its
initial position. In a region of homogeneous chemical composition this results in the
usual Schwarzschild criterion for instability:
\begin{equation}
\nabla_{\rm{rad}} > \nabla_{\rm{ad}},
\end{equation}
where $\nabla_{\rm{ad}}=(\partial \ln T / \partial \ln P)_{\rm{ad}}$ is the adiabatic
gradient and $\nabla_{\rm{rad}} = (\partial \ln T / \partial \ln P)_{\rm{rad}}$ is the 
same gradient assuming all the energy is carried by radiation (one usually uses the
diffusion approximation for this expression). This equation simply says that the
steepest stable gradient is the adiabatic one, and if a steeper gradient is required 
to carry the energy by radiation, then radiation will fail and buoyancy will develop.

This condition however ignores the possibility of a variation in the chemical
composition. Including this leads to the Ledoux criterion for instability:

\begin{equation}
\nabla_{\rm{rad}} > \nabla_{\rm{ad}} + \left(\frac{\varphi}{\delta}\right)  \nabla_{\rm{\mu}},
\end{equation}
where $\varphi = (\partial \ln
\rho / \partial \ln \mu)_{P,T}$, $\delta=-(\partial \ln \rho / \partial \ln
T)_{P,\mu}$ and  $\nabla_{\rm{\mu}} = (\partial \ln \mu / \partial \ln P).$
The essential feature here is the appearance of the term
involving the gradient of the molecular weight (the multiplicative terms 
$\phi$ and $\delta$ come from the thermodynamics and are of little importance 
for the present discussion; both are unity for a perfect gas without radiation
pressure).

For thermohaline mixing we need the Ledoux criterion to be
broken, but with the added condition that the molecular weight gradient decreases with depth, i.e.
\begin{equation}
 \nabla_{\rm{\mu}} < 0.
\end{equation}

Thermohaline mixing has a long history.
\citet{1969ApJ...157..673S} were the first to consider how thermohaline mixing may impact 
upon the structure and subsequent evolution of a star. They theorised that accreted 
material from a helium-rich companion would be thermohaline unstable and may be 
responsible for the pulsations of  $\beta$ Cepheids.

Soon after \citet{1970ApJ...162L.125A} and \citet{1971ApJ...168...57U} recognised
an unusual property of  the reaction
\begin{equation} \label{eqn:He3}
 \el{3}{He}\left( \el{3}{He},2 \rm{p}\right) \el{4}{He},
\end{equation}
which is that despite being a {\it fusion\/} reaction, it actually reduces the mean 
molecular weight because it produces three particles from two. (These particles also
have more kinetic energy than the initial two particles, because the reaction is exothermic.)
They thought the reaction may have an important role during pre-main-sequence contraction,
and could lead to the situation where $\nabla_{\mu} < 0$ and a 
thermohaline instability may develop. 
This was later found to have little effect due to the short time scale of the pre-main-sequence.

\citet{1972ApJ...172..165U}  was the first to derive an expression for the turbulent diffusivity 
in a perfect gas. He considered thermohaline mixing during the core flash where off-centre 
ignition leads to carbon-rich material sitting on material that is helium-rich.  
\citet{1980A&A....91..175K} extended this to allow for a non-perfect gas which 
included radiation pressure and degeneracy. Thermohaline mixing is unlikely to 
have an appreciable effect during the core flash because of the very short timescale 
of the flash. \citet{2006ApJ...639..405D}  showed 
that a small amount of overshooting inwards could remove the molecular weight
inversion on a dynamical timescale, thus removing the possibility of any thermohaline mixing.

\section{Thermohaline Mixing on the First Giant Branch}

A further application, and the one of interest to us here,  was proposed 
by EDL06. During main-sequence evolution a low mass star produces
a substantial amount of $\el{3}{He}$. 
Close to the centre all of the $\el{3}{He}$  has been destroyed in completing 
the pp chains. But at lower temperatures the $\el{3}{He}$ remains (e.g. see Fig 1 of EDL08).
When the FDU occurs it mixes the envelope abundances over a large
region, producing a homogeneous envelope with a $\el{3}{He}$ content that 
is far above the equilibrium value.
Following FDU, the convective envelope recedes, 
leaving behind a region that is homogeneous in composition.  

When the hydrogen burning shell advances, 
the fragile \el{3}{He}  begins to burn and the resulting decrease in the
molecular weight leads to an inversion developing. Note that \el{3}{He} 
burning alone is not sufficient to drive the mixing. Whenever there is hydrogen 
burning via the pp chains the \el{3}{He} + \el{3}{He} reaction (Equation \ref{eqn:He3}) will 
decrease $\mu$. However the combined effect of the other reactions and 
the low abundance of \el{3}{He} makes the effect of the \el{3}{He} + \el{3}{He}  reaction 
negligible. If one can homogenise the region beforehand, then the effect of \el{3}{He} + \el{3}{He}  
dominates. This way thermohaline mixing
is initiated at essentially the same luminosity as the bump in the luminosity function,
since both are caused by the H-shell reaching the abundance discontinuity left behind from FDU.

From equation (4) it can 
be shown that the change in $\mu$ must be (EDL08)
\begin{equation}
\delta\mu = \mu^2 \delta(X(\el{3}{He}))/6.
\end{equation}
and they  found 
the magnitude of such inversions to be  of the 
order ${\Delta \mu}/{\mu}$ $\sim$ 10$^{-4}$. Although this seems small, 
convection is in fact driven by a similarly small superadiabaticity. 

\indent 
We now try to understand what happens to the material in this
region of the star. When \el{3}{He} burns, a parcel forms that is 
slightly hotter and has lower molecular weight than its surroundings. 
This means it has a higher pressure than it requires for its position. Hence
it quickly expands (and begins to cool) in order to establish pressure equilibrium.
The expansion reduces the density and therefore the element becomes buoyant. 
The parcel rises until it finds an equilibrium point where the external 
pressure and density are equal to that inside the bubble. This is expected 
to be a small displacement which occurs on a dynamical
timescale. 

\indent 
As the molecular weight inside the bubble is lower than its surroundings 
the equilibrium point must correspond to a place where the external temperature 
is higher than that of the bubble. The temperature inside the bubble will 
be lower than its surroundings (we may assume  a perfect gas equation of state): 

\begin{equation} 
 \rm{ \frac{T_i}{T_o}=\frac{ \mu_i}{ \mu_o}},
\end{equation}
where subscript i denotes the inside of the bubble and subscript o denotes the
surroundings.
 As heat begins to diffuse into the parcel, we expect layers will start to strip
off in the form of long fingers. It is this
secondary mixing that governs the overall mixing timescale. The mixing
cycles in fresh \el{3}{He} from the envelope reservoir, to replace the
$\el{3}{He}$-depleted material that is rising toward the 
star's convective envelope. This upward-flowing
material will also have experienced other
burning, such as CN cycling, while in the hotter region. It will also
experience further burning in the future 
as the material is cycled through this region 
on subsequent occasions.

The diffusion of heat and composition is analogous to the situation 
in the oceans where haline fingers diffuse salt and heat, although in the 
stellar context we are dealing with a compressible flow. 
For a thorough investigation
of the differences between the oceanic and stellar case we refer to \citet{2010arXiv1006.5481D}.
The fact that in both instances there are two diffusive processes at work has 
seen the term `thermohaline' adopted in the astrophysics literature 
(e.g. CZ07a). 

As stated above, EDL08 originally referred to thermohaline
mixing on the RGB\footnote{Its application 
to the AGB is still contentious, see \citet{2010ApJ...713..374K}, \citet{2010MNRAS.403..505S}, \citet{2010A&A...521A...9C} and \citet{2010A&A...522A..10C}.} 
as ``$\delta \mu$ mixing'' to distinguish and emphasise the 
mechanism that drives the mixing, namely the slight difference in the molecular 
weight brought about by \el{3}{He} burning. Their simulations seemed to hint at the possibility of a dynamical phase in addition to the linear phase which CZ07a  modelled as UKRT thermohaline mixing.

Many groups have realised the importance of understanding the mixing and are investigating the instability that arises from the \el{3}{He} burning. The anlysis by \citet{2008ApJ...684..626D} of the mixing suggests a dynamical phase should not arise. Further complicating the picture are the 2D models by \citet{2010arXiv1006.5481D} that predict a different behaviour in the fluid compared to the 1D parameter fitting of CZ07a.  

We consider $\delta \mu$ mixing to be a strong extra-mixing 
candidate along the RGB as it can meet the  criteria outlined in the 
introduction (CZ07a, EDL08). To summarise, 
this \el{3}{He} burning drives mixing that begins after the hydrogen shell 
encounters the composition discontinuity left behind by FDU. We note that the depletion of
$\el{3}{He}$ is 
required to ensure that stellar nucleosynthesis and Big Bang Nucleosynthesis remain consistent 
\citep{1986ApJ...302...35D, 1995PhRvL..75.3977H, 1996ApJ...465..887D, 1999ApJ...510..217S, 2007A&A...467L..15C, 2007A&A...476L..29C}.  
Measurements  of  \el{3}{He} in HII regions and planetary nebula \citep{1984ApJ...280..629R,1995ApJ...441L..17H,1995ApJ...453L..41C,1998A&A...336..915C,1998SSRv...84..207T,2000A&A...355...69P,2003MNRAS.346..295R} match the predicted yields from Big Bang Nucleosynthesis. Canonical models predict that low-mass, main-sequence stars are net producers of \el{3}{He} which is returned to the ISM through mass loss. 
\citet{1995PhRvL..75.3977H} have shown that about 90\% of the \el{3}{He} produced on the main sequence 
must be destroyed to reconcile the two fields. We
saw above how these stars produced $\el{3}{He}$, and the newly discovered
mechanism now also destroys it, removing the inconsistency with Big Bang
Nucleosynthesis. This will still allow for the existence of the minority of planetary nebula 
that show large \el{3}{He} abundances 
\citep{1997ApJ...483..320B,1999ApJ...522L..73B,2006ApJ...640..360B, 2007A&A...476L..29C}. 

The thermohaline mixing mechanism described here will operate until 
the \el{3}{He} is destroyed, which may be beyond the tip of the RGB 
\citep{2010MNRAS.403..505S,2010A&A...521A...9C,2010A&A...522A..10C}. The molecular weight inversion caused by the \el{3}{He} burning 
creates an instability that cycles CN processed material into the envelope, having the desired 
effect on the surface abundances (CZ07a). Finally CZ07a, EDL08 and \cite{2010A&A...522A..10C},  demonstrated that the 
degree of mixing depends on mass and metallicity, most clearly seen in the 
different values of the carbon isotopic ratio in
Population I and Population II stars.

\section{The Formula for the Diffusion Coefficient}
It is common in stellar interior calculations to use a diffusion equation to simulate
mixing. It is important to remember that many mixing processes, such as convective
mixing,  are {\sl advective\/} in nature, not
{\it diffusive\/}. In the former, a property is distributed as a result of
bulk flows within the fluid. The latter follows from Fick's first law, which 
postulates that a flux exists between regions of high and low concentration of
the quantity of interest. The equation governing this is
\begin{equation}
\vec{j} = - D \vec{\nabla} n
\end{equation}
where $n$ is the density of the quantity and $\vec{j}$ is its flux. The parameter D is
the ``diffusion coefficient'' of the ``diffusivity''. 
For particles with a mean-free-path $l$ and typical speed $v$ then 
\begin{equation} \label{eq:diffeqn}
D =  {1\over 3} v l.
\end{equation}
As no theory for
time dependent mixing exists, it is common to use a diffusion equation to 
simulate mixing in stellar interiors. Historically, this is also how thermohaline mixing has been
calculated. A diffusion equation is solved for each species in the star, and
thus determines the radial composition variation.

\indent EDL08 used the following formula, based on estimates of the mixing velocity  
and the convective formalism in their evolution code:

\begin{equation}
 D=\left\{\begin{array}{ccc}\frac{\finv r^2}{\tn}\thn(\mu - \mu_{\rm
min})&\mbox{ if }
(k\ge k_{\rm min})\\\\0 &\mbox{ if }(k\le k_{\rm min}) , \end{array} \right.
\end{equation}
where $k$ is the mesh point number, counted outwards from the centre of
the model, $k_{\rm{min}}$ is the value of $k$ for which $\mu$ takes its minimum
value of $\mu_{\rm{min}}$, $r$ is the
radial coordinate,  $F_{\rm{inv}}$ is a constant  which is selected to obtain the
desired
mixing efficiency and $\tn$ is an estimate of the nuclear evolution timescale
(see EDL08).  This is an extension of the convective formula used in the
EDL08 calculations, with the addition of a $\mu - \mu_{\rm{min}}$ term to reflect the
driving by the $\mu$ inversion. Thus this formula is not a local one as the
value of $D$ at a point in the star depends on the value of $\mu_{\rm{min}}$ at some
other location in the star. Nevertheless, it is a phenomenological form that
recognises the role of the $\mu$ inversion in driving the mixing.

This formulation ensured the correct region was mixed but note that
the mixing speed is formally zero at the  position where $\mu$ has
its minimum even though
the mixing should presumably  be the most efficient at this point. EDL08 give upper
and lower estimates for the mixing velocity and find that they can change 
their free parameter  $F_{\rm{inv}}$, and consequently
alter the speed by three
orders of magnitude whilst still producing the
observed levels of \el{12}{C}/\el{13}{C} and leading to \el{3}{He} depletion. 

 CZ07a adopt the UKRT formulation for thermohaline 
mixing, which results from a linear analysis of the problem. They
cast it in the following way,
\begin{equation} \label{eq:kip}
D_t =  C_t \,  K  \left({\varphi \over \delta}\right){- \nabla_\mu \over
(\nabla_{\rm ad} - \nabla)} \quad \hbox{for} \;  \nabla_\mu < 0 ,
\label{dt}
\end{equation}
where  K is the thermal diffusivity and $C_t$ is a dimensionless free parameter.
In fact $C_t$ is related to the aspect ratio, $\alpha$, of the fingers in the following manner:

\begin{equation} \label{eq:ct}
C_t = {8 \over 3} \pi^2 \alpha^2
\end{equation}
 
The appropriate value to use for this parameter remains uncertain.
To understand thermohaline mixing fully clearly requires a hydrodynamic
theory that will determine the diffusion coefficient and any associated
parameters. In lieu of such a theory, we can compare with observations
and see what form of $D$ and values of constants are needed to match
the real world. This is the approach we adopt in this paper.

Alternatively, one could try to determine $C_t$ or $\alpha$ by comparison to 
laboratory experiments and the oceanic case, which has been well studied.
Stommel and Faller carried out experiments of fluids in laboratory conditions. 
This work, published in \citet{stern}, saw the development of long salt fingers 
with lengths that were larger than their diameters ($\alpha$ $\simeq$5). This led \citet{1972ApJ...172..165U}  to
determine that  $C_{\rm{t}}$  $\simeq$ 658 which was the basis of the choice by CZ07a to use $C_{\rm{t}} = 1000$. The huge differences between
incompressible salty water and the the self-gravitating plasma we are
considering make direct comparisons very difficult, however (see
\citealt{2010arXiv1006.5481D}). In contrast to the oceanic case, Kippenhahn 
envisaged the classical picture where mixing is due to blobs. If Kippenhahn's 
expression is cast into the same form as Equation \ref{eq:kip} then $C_{\rm{t}}$ $\simeq$ 12. This value seems to agree with the 2D-hydrodynamical models of \citet{2010arXiv1006.5481D} and 1D models of \citet{2010A&A...521A...9C}, but is inconsistent with the value preferred by CZ07a and the present work (see below).

\section{Models and Results}
 Messier 3 was chosen as a test case for thermohaline mixing in Population~II
stars because in many respects it can be considered a typical
globular cluster. A metallicity of [Fe/H] $= -1.4$ \citep{2004AJ....127.2162S, 2005AJ....129..303C}
means that it falls near the mode in the metallicity
distribution of halo globular clusters. Among clusters of similar
metallicity,
the horizontal branch morphology of M3 has stars on both the red and blue
sides
of the RR Lyrae gap \citep{1970ApJ...162..841S,1994A&A...290...69B}, and the colour
distribution is intermediate between extremes such as those of M13
and NGC 6752 (whose HBs have more extended blue tails; \citealt{1970ApJ...159..443N, 1980JKAS...13...15L}) and NGC 7006 (whose HB stars are more concentrated to the
red side
of the RR Lyrae gap; \citealt{1967ApJ...150..469S}). Red giants with enhanced
$\lambda$3883 CN bands in their spectra were discovered in M3 by Suntzeff
(1981), and further studied by \citet{1984ApJ...287..255N}. The CN distribution
contains both CN-strong and CN-weak giants (i.e., high and low nitrogen
abundance giants) as is typical of other clusters of similar metallicity
\citep{1984ApJ...287..255N}. The behaviour of the CN bands at a given luminosity
on the RGB appears to anticorrelate with carbon and oxygen abundance
\citep{1981ApJS...47....1S,1996AJ....112.1511S}. When giants of different luminosity are
compared a
significant trend becomes apparent in the carbon abundance. Among red
giants
with absolute magnitudes brighter than the horizontal branch [C/Fe]
declines
with increasing luminosity \citep{1981ApJS...47....1S,1989asgc.conf...71S,2002PASP..114.1097S} in a manner
that
is typical of other globular clusters and halo field giants \citep{2003PASP..115.1211S}. It is this behaviour of carbon in particular that is of great use
in
constraining models of thermohaline mixing. The carbon isotope ratio
$^{12}$C/$^{13}$C has a low value of $\sim$ 4-6 among the bright giants
\citep{2003AJ....125..794P,2003MNRAS.345..311P}, although the
number of stars for which this has been measured is small. In addition,
M3
displays O, Na, and Al abundance inhomogeneities of the type that are
commonplace within globular clusters of similar metallicity \citep{2005A&A...433..597C}. The cluster
has
a population of stars that are enhanced in Na and Al but depleted in
oxygen
\citep{1992AJ....104..645K, 2000AJ....120.1364C, 2004AJ....127.2162S, 2005AJ....129..303C, 2005PASP..117.1308J}, but without the extremes in O
abundance depletion and Na enhancement that are found in the oft-compared
cluster M13 \citep{2004AJ....127.2162S}. A substantial component of the
O-Na-Al variations in M3 are arguably the products of early cluster
self-enrichment, and thermohaline mixing on the first ascent giant branch
is not expected to have played a role in their production. Work by CZ07a and \citet{2010A&A...522A..10C} also show that these elements are not affected by thermohaline mixing.

\begin{figure} 
 \includegraphics[scale=0.35]{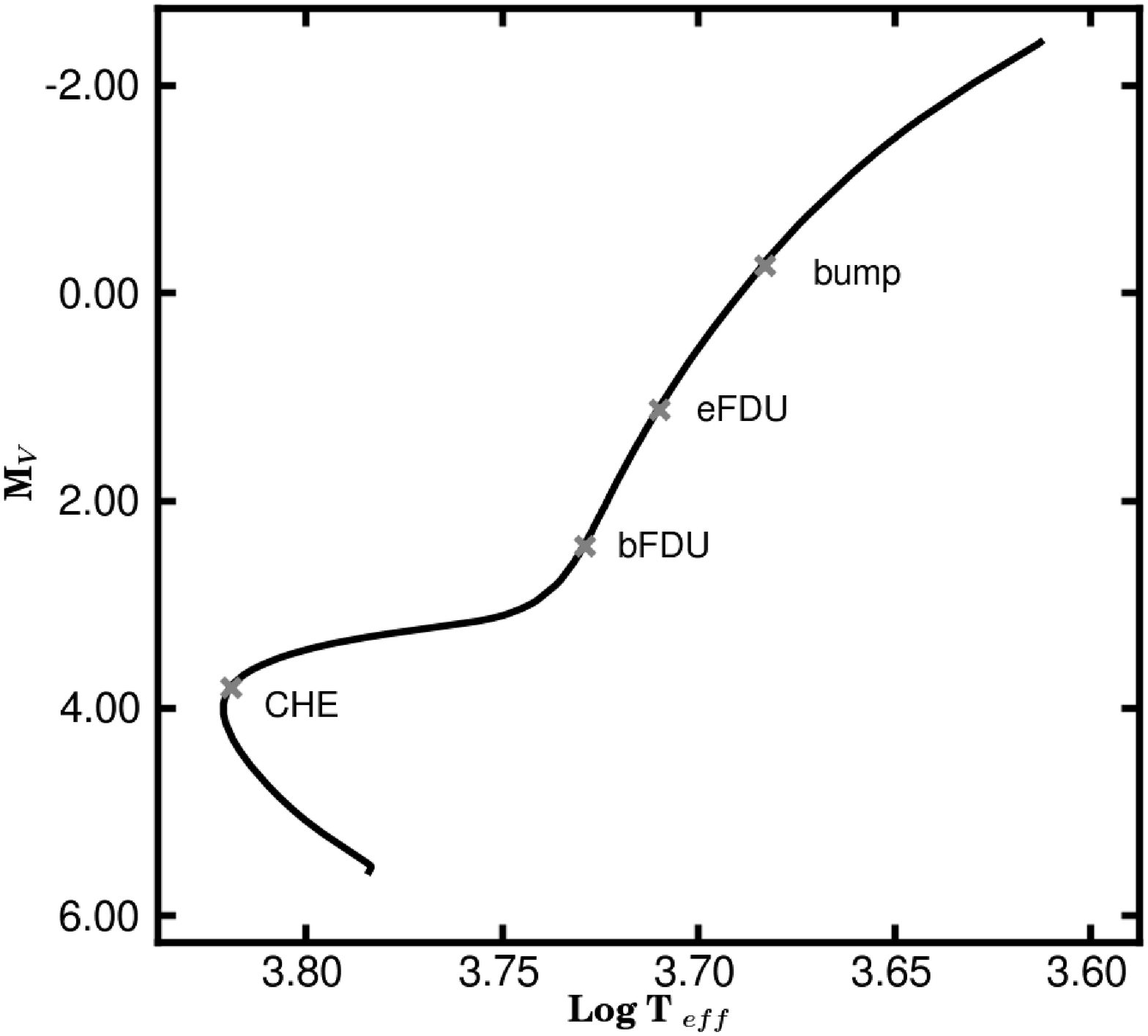}
\caption{HR Diagram for our star that models M3. We mark: core hydrogen exhaustion (CHE) and the beginning of FDU (bFDU) which we define as the point where the penetration of the envelope begins to affect the surface abundances. Also marked is the end of FDU (eFDU) and the bump, after which extra mixing is expected to begin.}
 \label{fig:evoln}
\end{figure}

\begin{figure} 
 \includegraphics[scale=0.28]{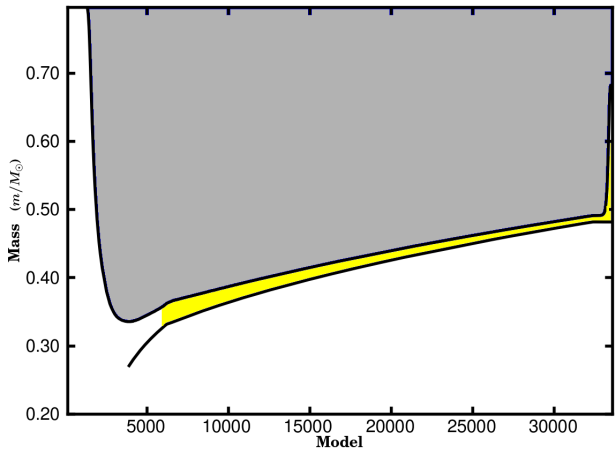}
\caption{A simplified Kippenhan plot for our model star. We plot the interior of the star in mass co-ordinates against model number which is a non-linear proxy for time. The grey shaded region shows the penetration of the convective envelope, its deepest point coinciding with the end of FDU. We also mark the top of the hydrogen-burning shell. Unshaded regions denote radiative zones, whilst the shaded region between the shell and the envelope represents the location where thermohaline mixing occurs in the star. Although not obvious on this scale there is still a small radiative buffer between the top of the shell and the thermohaline mixed regions. Thermohaline mixing has the effect of extending the convection zone and allows material to be mixed closer to the shell where it can be processed.} 
 \label{fig:Kippenhahplot}
\end{figure}

\begin{figure} 
 \includegraphics[scale=0.4]{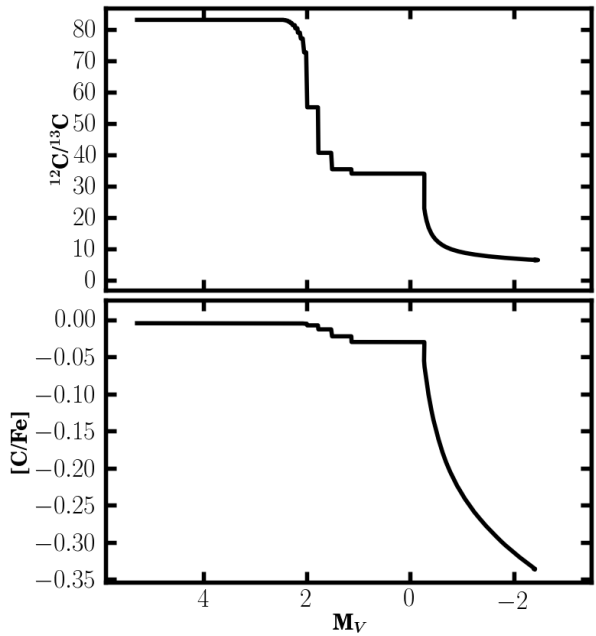}
\caption{Top panel: $^{12}$C/$^{13}$C as a function of magnitude. Bottom panel: [C/Fe] as a function of magnitude. In this figure we compare the evolution of  [C/Fe] to that of $^{12}$C/$^{13}$C. The UKRT thermohaline mixing formula with \textit{C$_{t}$}=1000 has been used in this model. It can be seen that the isotopic ratio reaches equilibrium soon after the onset of extra mixing, where as [C/Fe] depletes along the entire RGB.} 
 \label{fig:carbon}
\end{figure}

\begin{figure} 
 \includegraphics[scale=0.4]{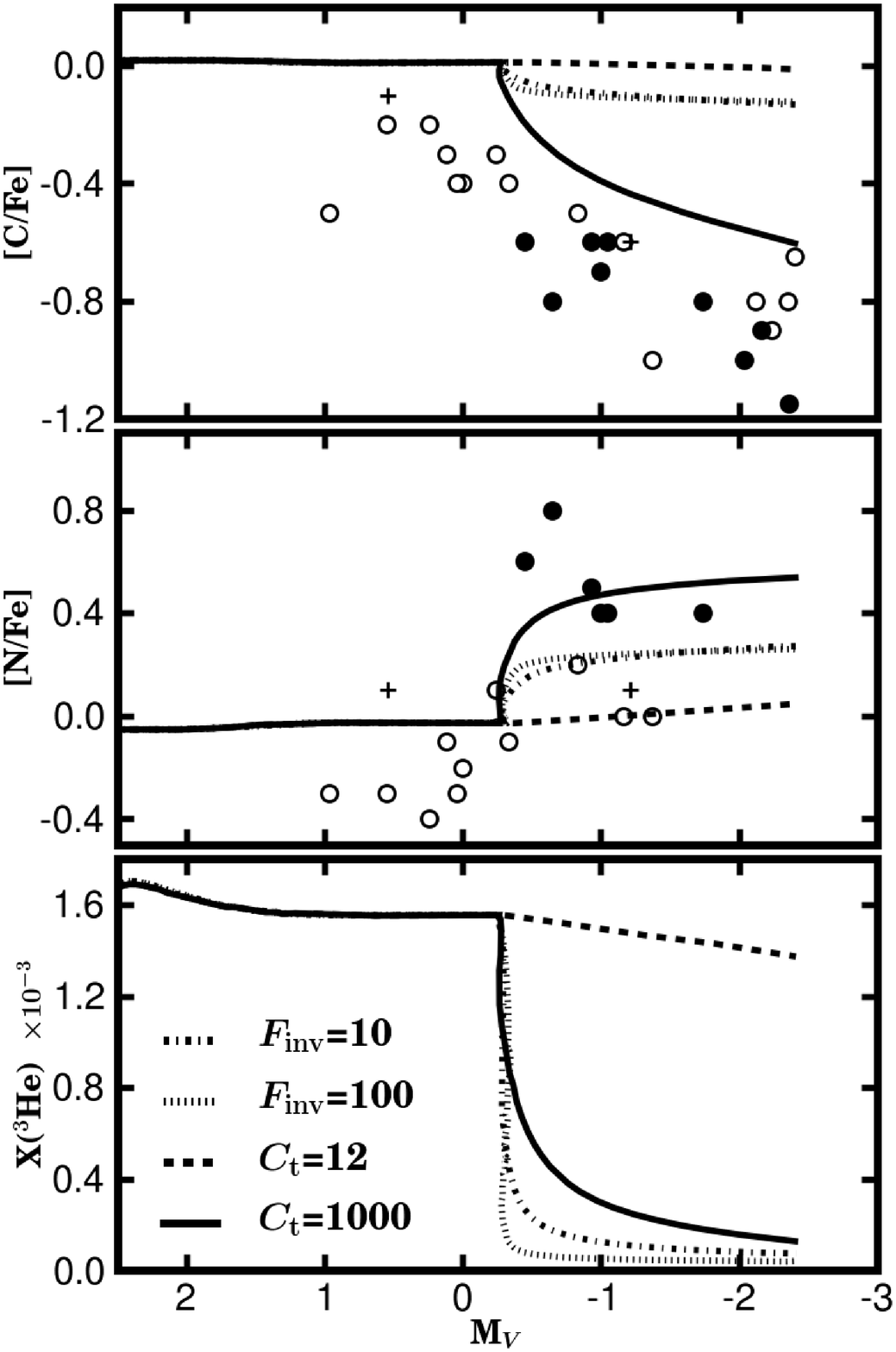}
\caption{Abundances as a function of visual magnitude. In this figure we 
compare the UKRT prescription for the diffusion coefficient to that of EDL. 
Open circles denote CN-weak stars in M3, filled circles denote CN-strong stars 
and crosses represent stars of intermediate CN strength, with all data taken from \citet{2002PASP..114.1097S}. 
The two UKRT models 
correspond to the dashed line for \textit{C$_{t}$}=12 and the solid line for 
\textit{C$_{t}$}=1000. The two EDL models are denoted by the dotted line for  
$F_{\rm{inv}}$=100 and the dot dashed line for  $F_{\rm{inv}}$=10, these sit 
nearly on top of each other on this scale. The models are all of mass 
$M$=0.8$\Mo$ and $Z$=5 $\times$ 10$^{-4}$ with solar scaled CNO abundances. The errors in magnitude are smaller than
the symbols used. The maximum error in [C/Fe] and [N/Fe] is $\pm 0.3$ dex. }
 \label{fig:kipedl}
\end{figure}

We use MONSTAR (The Monash version of
the Mt.Stromlo evolution code; see \citealt{2008A&A...490..769C}) to 
produce stellar models for M3. The stars are evolved from the zero-age 
main-sequence until the core flash. They are of mass M=0.8 $\Mo$, 
metallicity Z=0.0005 and have a solar-scaled CNO abundance, as a first approximation.
Population II subdwarfs in the globular cluster metallicity range         
tend to have [C/Fe] $\sim  0$ (e.g. \citealt{1985ApJ...289..556L}). In M13 the work of \citet{2004AJ....127.1579B} shows an upper envelope to the data 
that approaches [C/Fe] $\sim  0$ to $-0.2$ near the
main sequence.

Our mixing length parameter $\alpha$ is set to 1.75 and we run without any overshoot.
Mixing is calculated using a diffusion equation, and we
investigate different 
formulae for the diffusion co-efficient used for thermohaline mixing.
These parameters approximate the stars in M3 quite well: M3 has an  age
estimated by various authors to be
between 11.3 to 14.2 Gyr \citep{1992ApJ...394..515C,1996MNRAS.282..926J,2000ApJS..129..315V,2002A&A...388..492S,2004AAS...205.5301A}
and [Fe/H]=$-$1.5. Figure \ref{fig:evoln} shows the evolution of our model star in the
HR Diagram. We mark the end of core H exhaustion (CHE in the figure, at an
age of t$=$12.5 Gyr), the beginning and end of FDU (bFDU and eFDU respectively)
as well as the position of the LF bump.
Figure \ref{fig:Kippenhahplot} shows a Kippenhahn diagram for the evolution 
of our star. Here we see the convective envelope moving inwards (coinciding with the beginning of FDU), whilst the deepest point of penetration marks the end of FDU. The location of thermohaline mixing is the lightly shaded region between the envelope and the advancing hydrogen burning shell. In Figure~\ref{fig:carbon}, using the UKRT formulation for thermohaline mixing and \textit{C$_{t}$}=1000, we
plot the variation of the surface [C/Fe]  as well as the $^{12}$C/$^{13}$C ratio
versus $M_V$. Here it can clearly be seen that the $^{12}$C/$^{13}$C ratio drops below 10 very quickly, finally reaching as low
as 6.5. This is in good agreement with the observations of 4-6 by \citet{2003AJ....125..794P} and \citet{2003MNRAS.345..311P}. The gradual depletion of carbon throughout the entire RGB on the other hand provides a more sensitive constraint on the mixing.

Figure \ref{fig:kipedl} compares the results with the EDL08 and UKRT formulae 
for the diffusion coefficient. In this figure open circles denote CN-weak stars, 
filled circles denote CN-strong stars and crosses represent 
CN-intermediate stars all from M3\footnote{We note that there appears to be a lack of CN-strong stars on the lower RGB. This is an artifice of the original \citet{1981ApJS...47....1S} study in which the lower luminosity stars that were observed happened to be CN-weak. \citet{1984ApJ...287..255N} showed that CN-strong giants do exist in M3 at luminosities corresponding to the faint limit of the \citet{1981ApJS...47....1S} survey. However, their later study concentrated on the $\lambda$3883 CN band strength and did not measure the [C/Fe] or [N/Fe] abundances. Consequently, absence of CN-strong stars at the faintest limits of the abundance plots in our paper is a consequence of observational effects.}. In this figure as well as the proceeding figures the errors in magnitude are smaller than the symbols used. The random errors in the abundance measurements can be up to $\pm 0.3$ dex according to \citet{1981ApJS...47....1S}, although this does not take into account potential systematic errors due to limitations in the stellar atmospheres code used in the abundance derivations.  We have run four models to compare to 
the observations, two with thermohaline mixing that use the UKRT formula 
for the diffusion coefficient and two models that use the EDL08 formula.  
The dashed line and the solid line are variations of the UKRT prescription, 
the broken line uses  \textit{C$_{t}$}=12 ($\alpha \sim 1$) as suggested by \citet{1980A&A....91..175K}  
while the model represented by the solid line uses \textit{C$_{t}$}=1000 ($\alpha \sim 6$),  
a geometry more akin to CZ07a. The dotted and
dot-dashed lines (which nearly sit on top of each other) are EDL08 models 
where $F_{\rm{inv}}$=100 corresponds to the dotted line and  
$F_{\rm{inv}}$=10 corresponds to the dot-dashed line. 
The two choices  of  $F_{\rm{inv}}$ lie within the three orders of magnitude 
that lead to the required level of \el{12}{C}/\el{13}{C} depletion as discussed in EDL08. 

\begin{figure} 
 \includegraphics[scale=0.4]{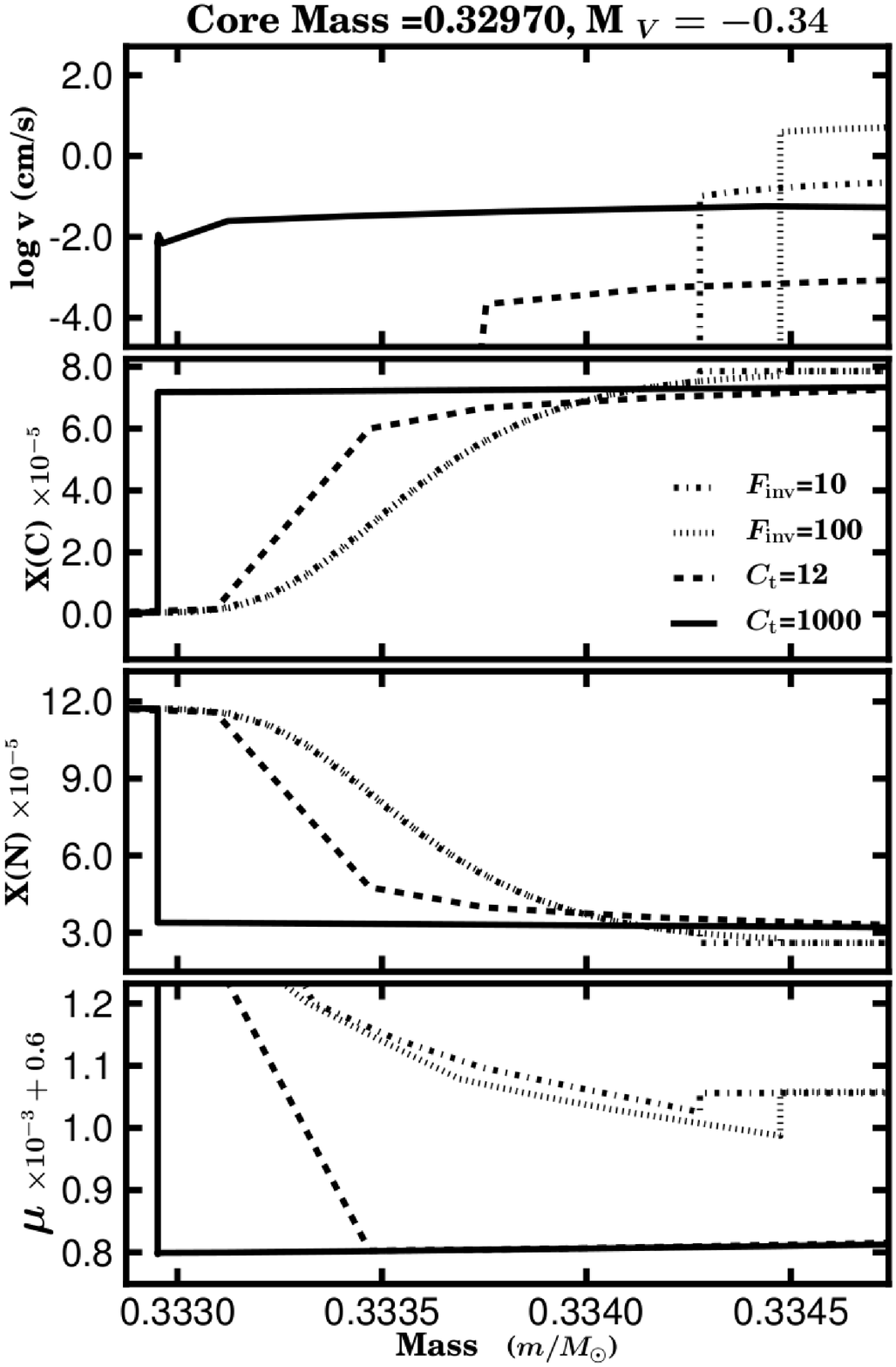}
\caption{Various profiles for the same models as Figure \ref{fig:kipedl}. 
The same symbols are used. The profiles are all taken at the same stage in 
evolution, that is when the hydrogen-exhausted core mass $M_{\rm{c}}$=0.32970 $\Mo$ and at and $M_V = -0.34$.  
Top Panel: the mixing velocity in the region above the shell. Second Panel: the 
carbon abundance in the region above the shell. Third Panel: the nitrogen abundance 
in the region above the shell.  Bottom Panel: highlights the respective locations 
where the minimum value of $\mu$ occurs.}    
 \label{fig:kipedlprofile}
\end{figure}

The two EDL08 models destroy the available \el{3}{He} very quickly and are 
unable to cycle much CN processed material to the envelope.  As previously 
mentioned, in the EDL08 formulation of the diffusion coefficient the mixing 
is dependent on the difference in $\mu$, i.e. $(\mu-\mu_{\rm{min}})$. 
Material near the envelope is cycled down faster and the mixing becomes 
less efficient closer to the location where $\mu$ has its minimum. Material 
is not replenished at the same rate it would be with a local condition on 
the $\mu$ gradient as seen with the UKRT prescription. This in turn affects 
the depth at which the minimum value of $\mu$ occurs. We can see in 
Figure \ref{fig:kipedlprofile} that the location at which $\mu$ has its 
minimum occurs further out in the EDL08 scheme. The steep temperature gradient 
ensures the fragile \el{3}{He} is easily destroyed while the carbon-rich 
material is not exposed to temperatures where it can burn significantly. 
CN cycling is therefore reduced whilst the driving fuel, \el{3}{He}, is 
easily processed irrespective of one's choice of  $F_{\rm{inv}}$. The 
bottom panel in Figure \ref{fig:kipedlprofile} clearly demonstrates the 
dependence of the location of the minimum value of $\mu$ on the mixing scheme. 
The four models in Figure \ref{fig:kipedl} are plotted with the same symbols 
in Figure \ref{fig:kipedlprofile}. We have plotted the velocities, carbon, 
nitrogen and molecular weight profiles at a hydrogen-exhausted core mass 
of M$_c$=0.32970 $\Mo$ corresponding to $M_V = -0.34$ . The bottom panel demonstrates the depth to which the 
 models mix. This highlights the fact that the two UKRT models 
have their $\mu$ minimum occurring closer to the hydrogen shell and mix 
deeper than the EDL08 models at the same core mass. As expected the two 
EDL08 abundance profiles show little CN processing. As shown by EDL08 this 
is sufficient to produce the observed  \el{12}{C}/\el{13}{C} values 
but as Figure \ref{fig:kipedlprofile} shows it is not enough to reproduce the [C/Fe] or [N/Fe] variations. Note that to calculate a velocity from Equation \ref{eq:diffeqn} we require a value for $l$. We have used $l = 1.75 \times$ H$_p$ where H$_p$ is the pressure scale height. This may not be an accurate estimate but it is expeted to give us an indicative velocity.

\begin{figure} 
 \includegraphics[scale=0.4]{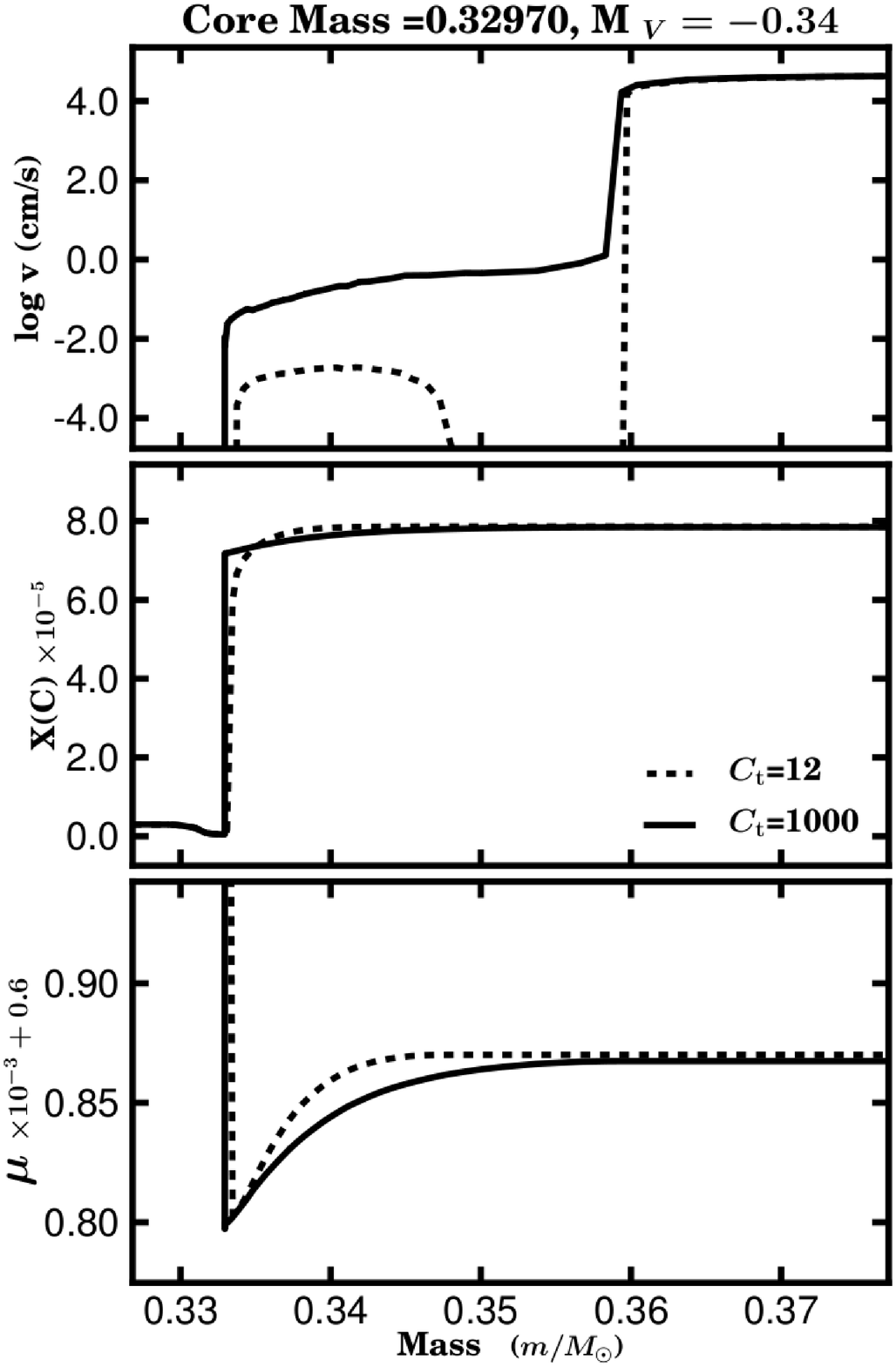}
\caption{Velocity, $X_{\rm{c}}$ and $\mu$ profiles for the two UKRT models. 
This is an expanded plot of the first panel in Figure \ref{fig:kipedlprofile} 
so is once again taken at a hydrogen-exhausted core mass of M$_c$=0.32970 and $M_V = -0.34$ . 
The \textit{C$_{t}$}=12 model (dashed line) mixes two orders of magnitude 
more slowly than the \textit{C$_{t}$}=1000 model (solid line). It also fails 
to transport material to and from the envelope at this stage. It does connect 
with the envelope before the end of the RGB.}
 \label{fig:kipvel}
\end{figure}

The surface abundances are determined by both the depth to which 
the material is mixed as well as the speed of mixing. The depth of 
mixing in a ``steady state'' is determined by the location of the 
$\mu$ minimum which is in turn dependent on the mixing speed, so 
that the two are not independent. Consider the case of very fast mixing, 
such as convection. The abundance would be homogeneous from the 
deepest point of mixing  to the surface. Clearly the surface abundance 
is controlled by the burning conditions at the bottom of the mixed 
region - a case we call ``burning limited.'' Now consider the case 
of very slow mixing. The abundance profile is marginally altered from 
the radiative case. Although the material is mixed to the surface it 
proceeds so slowly that the advance of the burning region is almost 
independent of the (slow) mixing - just as it would be in the 
limiting case of no mixing (i.e. a radiative zone). Hence in the 
slow mixing case the surface abundances are ``transport limited.'' 
In Figure \ref{fig:kipedl} the bottom panel suggests that the UKRT 
\textit{C$_{t}$}=12 model is transport limited and this, not the 
depth to which material is mixed, is responsible for the lack of 
processing. The bottom panel in Figure \ref{fig:kipedlprofile} shows 
that the EDL08 models do not mix as deep as the UKRT models. Yet the 
EDL08 models destroy essentially all of the \el{3}{He} (see Figure \ref{fig:kipedl}) 
while this is not true for the UKRT formulation. \el{3}{He} and 
consequently carbon in the \textit{C$_{t}$}=12 model are not destroyed 
because not enough material is physically transported through the 
radiative region and processed before the end of the RGB. 
In Figure \ref{fig:kipvel} we see that there is  is almost a difference 
of two orders of magnitude between the mixing speeds in the two UKRT 
models. In addition at this stage of the evolution the \textit{C$_{t}$}=12 model 
does not connect the region of the $\mu$-inversion (where the burning occurs) 
with the envelope to cycle in fresh fuel (it does so further up the RGB): this 
further hinders any processing.            

It appears that the requirement for significant carbon depletion is two 
fold: the position of the minimum in $\mu$ must occur at a reasonably 
high temperature and we need a sufficiently fast mixing velocity. As a 
result of our comparison with M3 we prefer the UKRT formalism over 
that used by EDL08.

\section{Constraining the Diffusion Coefficient}
\begin{figure*} 
 \includegraphics[scale=0.3]{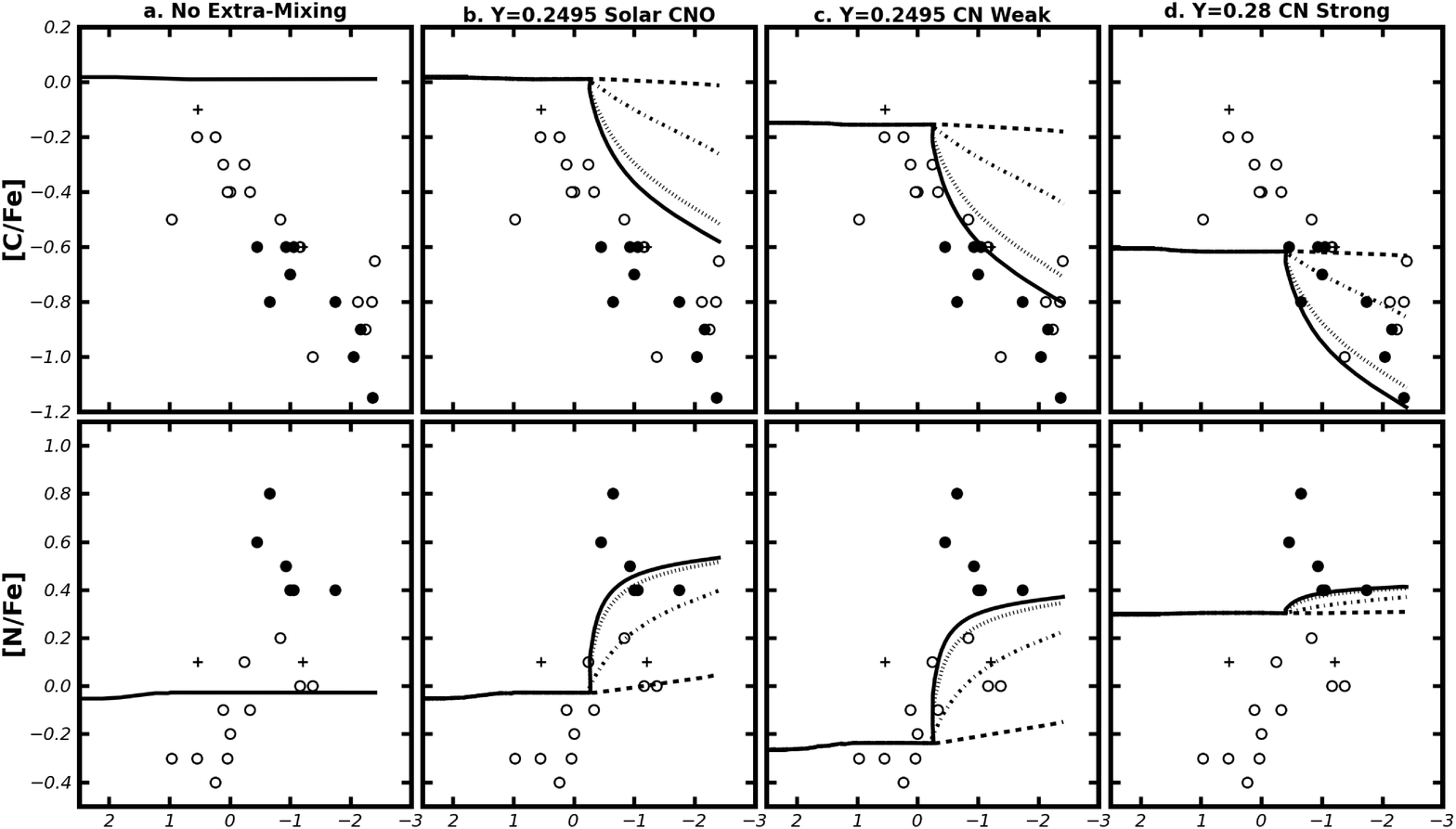}
\begin{center}
\ \ \ \ \ \textbf{M\boldmath$_V$}
\end{center}
\caption{Carbon and nitrogen abundances as a function of absolute 
magnitude in M3. Here we compare our models to the \citet{2002PASP..114.1097S} 
compilation of M3 observations. Open circles denote CN-weak stars, 
filled circles  CN-strong stars, whilst the crosses represent stars of 
intermediate CN strength. The errors in magnitude are smaller than
the symbols used. The maximum error in [C/Fe] and [N/Fe] is $\pm 0.3$ dex. Where extra mixing is included in our models 
dashed lines correspond to \textit{C$_{t}$}=12 ($\alpha \sim 1$), dot-dashed lines to 
\textit{C$_{t}$}=120 ($\alpha \sim 2$), dotted lines to  \textit{C$_{t}$}=600 ($\alpha \sim 5$) and solid 
lines to \textit{C$_{t}$}=1000 ($\alpha \sim 2$).  Figure \ref{fig:kipplot}c includes 
models with the following initial CNO abundances X(C)=5.45$\times$10$^{-5}$, 
X(N)=1.5$\times$10$^{-5}$ and X(O)=2.86$\times$10$^{-4}$. In Figure 
\ref{fig:kipplot}d the initial CNO abundances are X(C)=1.9$\times$10$^{-5}$, 
X(N)=5.5$\times$10$^{-5}$ and X(O)=2.6$\times$10$^{-4}$.}
\label{fig:kipplot}
\end{figure*}

 As previously mentioned \citet{1980A&A....91..175K}, \citet{1972ApJ...172..165U} 
and CZ07a differ in their choice of $\alpha$, the aspect ratio of the fingers by a factor of five and this translates to a difference of two orders of magnitude in \textit{C$_{t}$}. We can also determine
a velocity for the mixing through a diffusion
equation.

Having selected our preferred formalism we can now use observations to constrain the diffusion coefficient.
In Figure \ref{fig:kipplot}a we illustrate the effect of standard evolution 
on the carbon and nitrogen abundances. In all the Figure \ref{fig:kipplot} 
panels, open circles denote CN-weak stars , filled circles denote 
CN-strong stars and crosses represent CN-intermediate stars as before. The uncertainty in the data is as specified in Section 5.  Figures 
\ref{fig:kipplot}b, \ref{fig:kipplot}c and \ref{fig:kipplot}d include 
thermohaline mixing with the UKRT prescription. In each panel we provide 
four models, these represent different values of \textit{C$_{t}$}.  
We include: \textit{C$_{t}$}= 12  as per \citet{1980A&A....91..175K} (dashed lines, $\alpha \sim 1$),  
\textit{C$_{t}$} =1000 as per CZ07a (solid lines, $\alpha \sim 6$) 
and  two intermediate values, \textit{C$_{t}$}=120 (dot-dashed lines, $\alpha \sim 2$) 
and \textit{C$_{t}$}=600 (dotted lines, comparable to Ulrich 1972 \textit{C$_{t}$}=658, $\alpha \sim 5$).  

The canonical evolution in Figure \ref{fig:kipplot}a shows very 
little carbon depletion following FDU. This is contrary to the 
observations for the upper part of the RGB. We include this panel to highlight to the reader 
the need for extra mixing. In Figure \ref{fig:kipplot}b we have assumed 
a scaled solar CNO abundance and varied the free parameter as described 
above. It is unlikely that M3 possesses a scaled solar abundance. 
\citet{1992ApJ...384..508R}  have already shown that Large Magellanic 
Cloud abundances are not scaled Solar and we would not expect the 
environment during globular cluster formation to resemble the solar neighbourhood. 

In Figure \ref{fig:kipplot}c we have altered the initial CNO abundances whilst 
keeping the metallicity and total CNO content constant. As the carbon abundance 
does not change a great deal before the onset of extra mixing we have assumed 
the most carbon-rich stars are representative of the pre-FDU abundance, 
similarly for the most nitrogen-poor stars. To match the constraints, 
the following initial CNO abundances were assumed X(C)=5.45$\times$10$^{-5}$, 
X(N)=1.5$\times$10$^{-5}$ and X(O)=2.86$\times$10$^{-4}$. We altered the 
carbon and nitrogen so that we arrived at our desired post-FDU value and 
adjusted the oxygen abundance in order to keep the total CNO 
constant\footnote{Note that the effect of FDU is very small so the pre-FDU 
and post-FDU values are nearly the same.}. We have decreased the carbon 
by a factor of 1.47 and the nitrogen by a factor of 1.63 (in relation to solar). 
These values could easily be 
fine-tuned to better fit the data but it is the proof of concept we are 
concerned with. The two slow mixing cases are unable to account for the 
depletion in carbon. Both \textit{C$_{t}$}=600 and \textit{C$_{t}$}=1000 
are a good fit to the carbon and the nitrogen. It does appear that the 
mixing should begin at a slightly lower luminosity, however
\citet{1978IAUS...80..333S} demonstrated that lowering the \el{4}{He} 
abundance will cause the extra mixing to begin earlier because the shell 
reaches the discontinuity sooner.  We find that our models
provide a fit to the upper envelope of the run of [C/Fe] vs $M_V$
in Figure \ref{fig:kipplot}c but can only fit the nitrogen for 
the CN-weak stars. This argues that much of the
spread in [N/Fe] in the lower panels of Figure 4 is dominated
by
a variation that was present among the cluster stars before
they
commenced red giant branch evolution. The fact that none of
the
models in the the lower panels of Figure 4 can reproduce the
observed spread in [N/Fe] is suggestive of the CN
inhomogeneity
in M3 having a predominantly primordial origin. This
inference
is consistent with the discovery in a number of clusters,
including M3 \citep{1984ApJ...287..255N}, that N abundance
variations
(or equivalently CN band strength variations) are found both
at
and below the magnitude of the bump in the RGB luminosity
function
(e.g., see \citealt{1991ApJ...381..160S,2004AJ....127.1579B,2005A&A...433..597C}; for the clusters NGC 6752 and M13 which
have similar [Fe/H] to M3).
This argues strongly for the spread being present in 
the stars at their birth. 

In Figure \ref{fig:kipplot}d we have assumed the CN-strong stars are a  
separate population with their own CNO abundance. Moreover we have 
assumed there is a CN-[C/Fe] anticorrelation in which the CN-strong 
stars are initially 0.4 dex depleted in carbon relative to the 
CN-weak stars \citep[and references therein]{2002PASP..114.1097S}.
The CN-strong stars
show the results of hot hydrogen burning in the gas from
which they formed, including 
CN and ON cycling. We have therefore changed the initial 
helium abundance from Y=0.2495 to Y=0.28 as well as the CNO values\footnote{We note that this is consistent with the growing literature on multiple stellar populations in GC and their presumed \el{4}{He} content \citep[and references there in]{2009IAUS..258..233P}.}. 
Here we set  X(C)=1.9$\times$10$^{-5}$, X(N)=5.5$\times$10$^{-5}$ and 
X(O)=2.6$\times$10$^{-4}$. We find in this case the slower mixing is 
a good fit for the carbon abundance whilst the faster mixing is able 
to account for the more extreme observations of carbon depletion. 
These models are unable to match the more extreme nitrogen enhancements 
in the CN strong stars.  The [N/Fe] is initially so high even on the 
main sequence that any enhancement due to extra mixing only slightly 
raises the [N/Fe]. We conclude that  much of the nitrogen spread in the CN 
strong stars is due to primordial variations as a result of hot 
hydrogen burning (in an early
generation of stars that instigated cluster enrichment) rather
than thermohaline mixing in the present-day giants.

\begin{figure} 
 \includegraphics[scale=0.4]{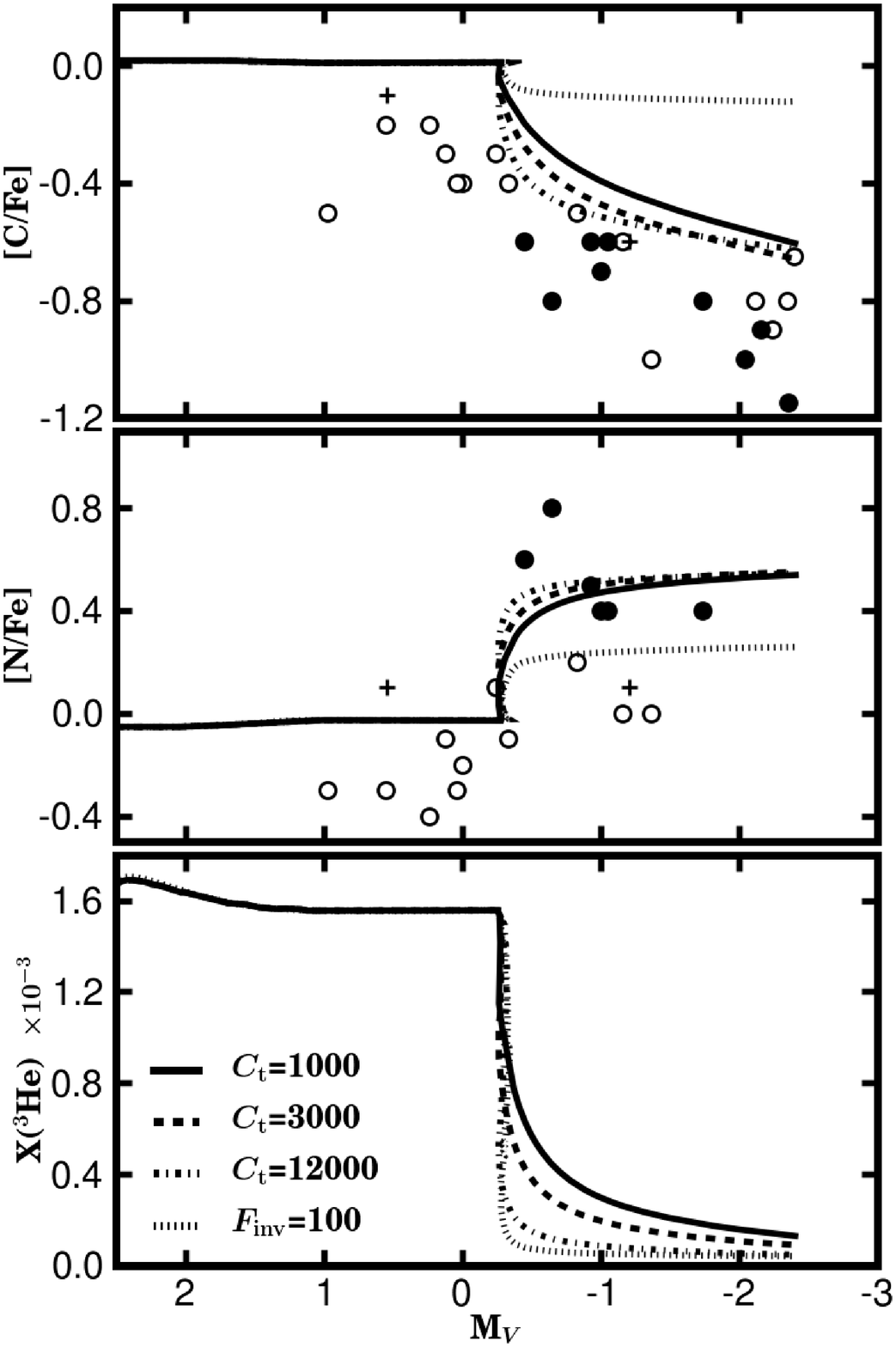}
\caption{Abundances as function of magnitude. We consider values 
of \textit{C$_{t}$} beyond 1000. M3 data is represented with the same 
symbols as in the previous figures. The dotted line corresponds to the EDL 
prescription with  $F_{\rm{inv}}$=100, the solid line corresponds to 
\textit{C$_{t}$}=1000, the dashed line to \textit{C$_{t}$}=3000  ($\alpha \sim$10) 
and the dash-dotted line to \textit{C$_{t}$}=12000 ($\alpha \sim$20). There 
is little difference in the level of carbon depletion once \textit{C$_{t}$} $>$ 1000,  
because the depletion is limited by the burning rate rather than the transport rate.} \label{fig:fastmix}
\end{figure}

Our findings are consistent with those of CZ07a in 
that \textit{C$_{t}$}=1000 is our preferred value of the free parameter 
in the UKRT thermohaline mixing prescription. In Figure \ref{fig:fastmix}  
we investigate the effect of increasing \textit{C$_{t}$} beyond 1000. 
The dotted line corresponds to the EDL08 prescription with  $F_{\rm{inv}}$=100, 
the solid line corresponds to \textit{C$_{t}$}=1000, the dashed line 
to \textit{C$_{t}$}=3000  ($\alpha \sim$10) and the dotted line 
\textit{C$_{t}$}=12000 ($\alpha \sim$20). Once again we use the same 
symbols for the M3 data. The initial abundances are solar scaled CNO, 
however it is the shape of the profiles we are interested in here. We 
include the EDL08 model to compare the similarities in the \el{3}{He} 
depletion as a function of magnitude. The fact that the UKRT \el{3}{He} 
profiles begin to approach that of the faster mixing EDL08 model once again 
suggests that depth to which the UKRT models mix is responsible for the 
carbon depletion. Initially the faster mixing cases are able to deplete 
the carbon more efficiently than our preferred value of \textit{C$_{t}$}=1000. 
Although all three values of \textit{C$_{t}$}  begin to deplete at different 
rates they eventually lead to similar levels of CN processing. The similar 
levels of carbon depletion suggest that a steady state has been reached and 
for \textit{C$_{t}$} $\geq$ 1000 the process is burning limited.

\section{Mass Loss}
\begin{figure}
 \includegraphics[scale=0.4]{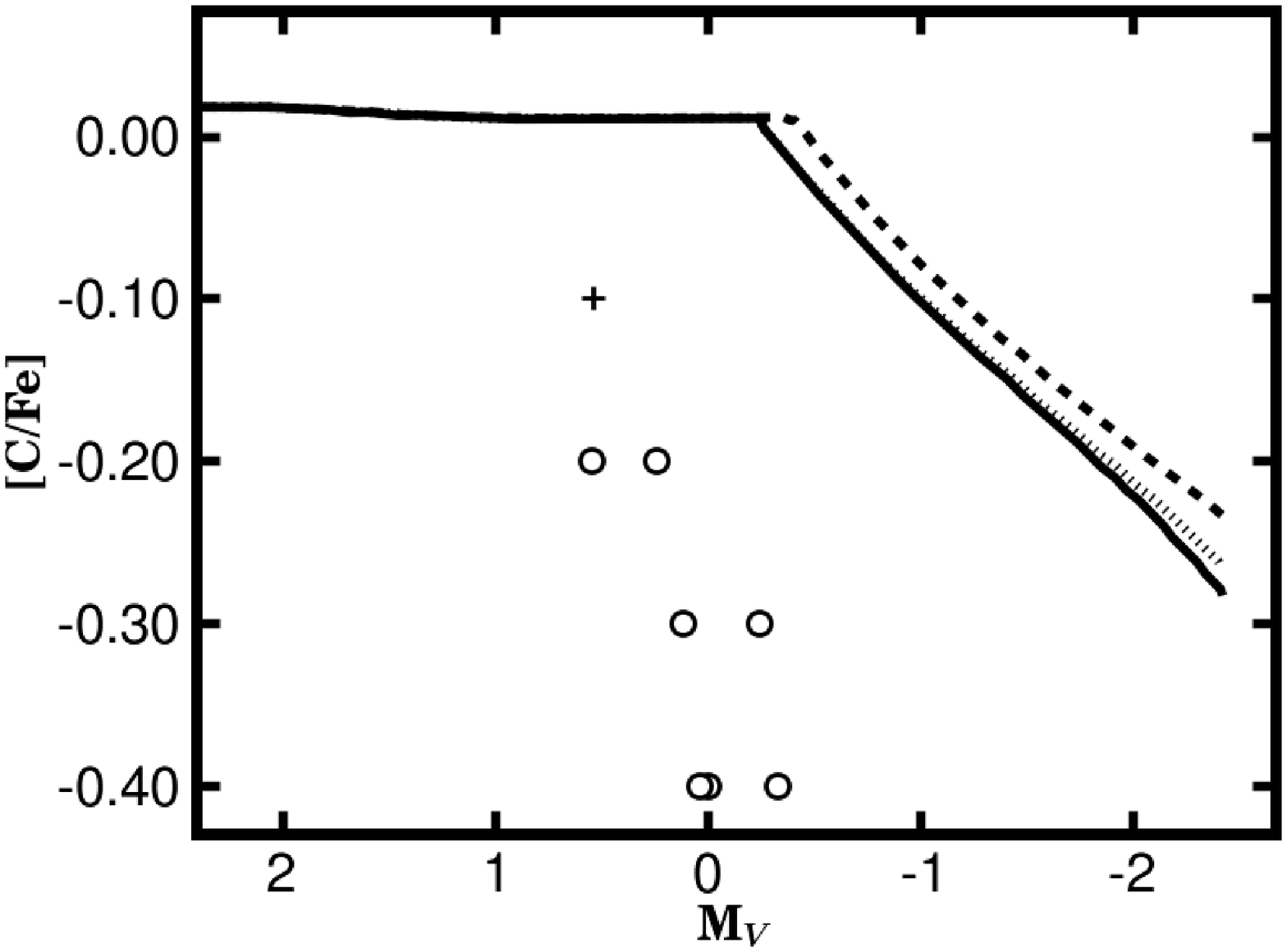}
\caption{Carbon abundance relative to solar as a function of magnitude. 
Here we plot UKRT \textit{C$_{t}$}=120 models with varying levels of mass 
loss. Symbols are described in previous figures. The dashed line corresponds to a model with no mass loss, the dotted 
line to standard Reimers' mass-loss rate and the solid curve twice the 
mass-loss rate. There is little effect on the carbon depletion.}
 \label{fig:massloss}
\end{figure}

Mass loss could have a significant effect on the level of carbon 
depletion in the models. If there is less envelope, the amount of 
carbon that needs to be processed to reduce the [C/Fe] is also less. 
In Figure \ref{fig:massloss} we investigate the role of mass loss 
in the UKRT formula for mixing.  We provide three models for the 
\textit{C$_{t}$}=120   case, standard Reimers' mass loss (dotted line), 
no mass loss (dashed line) and a factor of two increase on the 
standard mass-loss rate (solid line).
The effect of mass loss on the
surface  composition is greatest towards the tip of the giant branch where the
mass-loss rate is highest, but even there the alteration in [C/Fe] is
slight compared to a zero mass-loss model. These findings are consistent with EDL08, who also find the effect of mass loss
negligible after the onset of thermohaline mixing.

\section{The Onset of Mixing and the Location of the LF Bump}
\begin{figure}
 \includegraphics[scale=0.4]{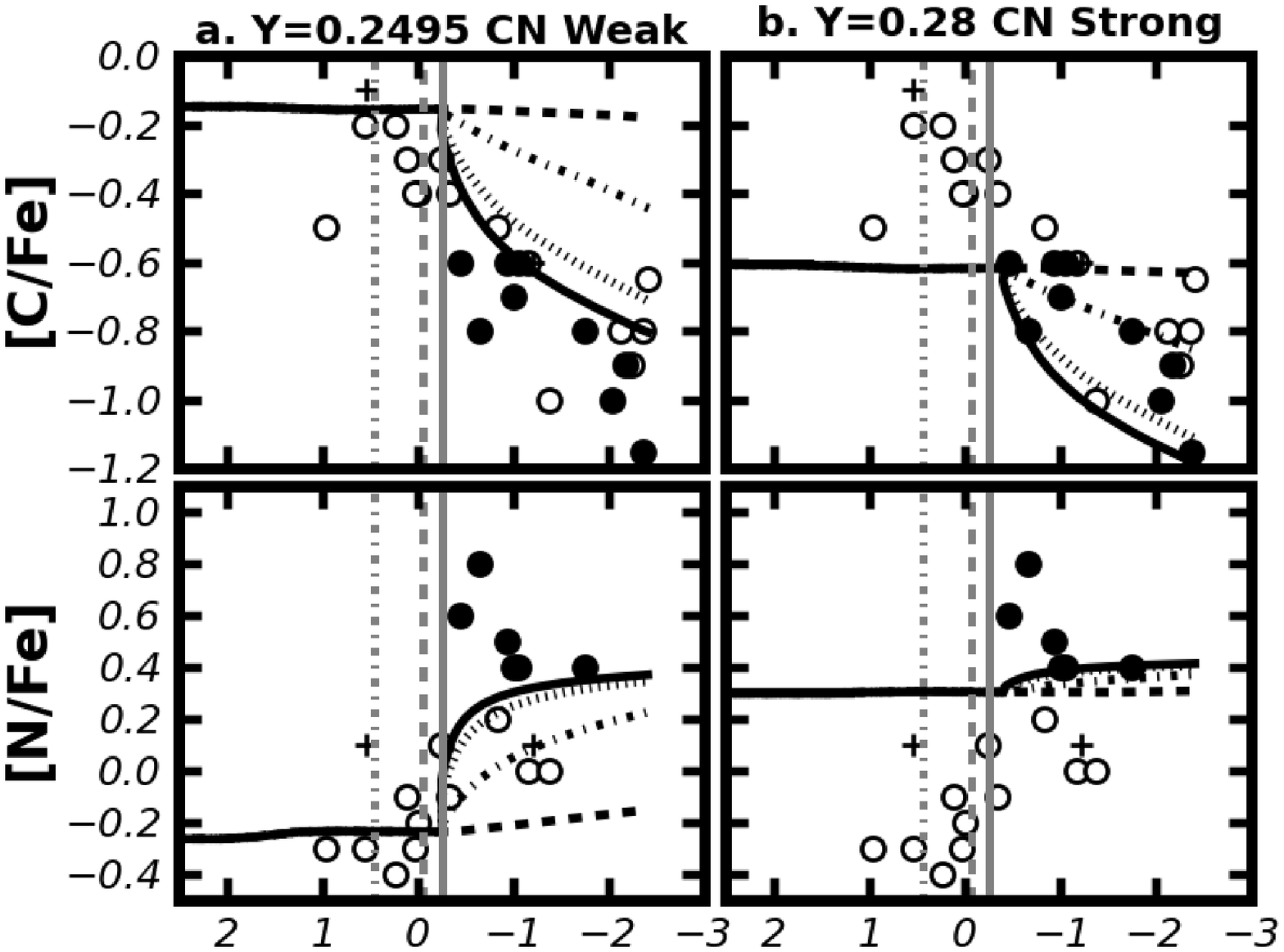}

\hspace{4cm} \textbf{M\boldmath$_V$}
\caption{Carbon and nitrogen abundances as a function of absolute 
magnitude in M3. These are the same panels as Figures \ref{fig:kipplot}c 
and \ref{fig:kipplot}d with the location where the local luminosity 
maximum occurs when the hydrogen shell meets the composition 
discontinuity (solid vertical line), the location of the LF bump 
for M3 from \citet{1990A&A...238...95F} (dashed vertical line) and 
location of the LF bump for M3 from \citet{2003PASP..115.1211S} 
(dot-dashed vertical line). }
 \label{fig:lfbump}
\end{figure}

Our final observational test concerns the onset of mixing and the 
observed location of the abundance changes. The hydrogen-burning 
shell will reach the composition discontinuity independent of any 
thermohaline mixing, which only begins after this event. Once the mixing
starts, the rate is dependent on the parameter $C_t$ and so
the surface abundance begins changing at a visual magnitude that 
depends slightly on $C_t$. 
The models for the four different values of $C_t$
in Figure \ref{fig:kipplot} commence thermohaline mixing within
$\Delta M_V =0.02$ mag of each other.
In Figure \ref{fig:lfbump} 
we have taken Figures \ref{fig:kipplot}c and \ref{fig:kipplot}d 
and marked the position of M3's LF bump according to \citep[grey 
dashed line]{1990A&A...238...95F} and the position according to 
\citet[dot-dashed grey line]{2003PASP..115.1211S}.  In all four 
panels we also include  the magnitude at which the hydrogen shell 
reaches the discontinuity in the CN weak models\footnote{Technically, we 
plot the local luminosity maximum that occurs during this phase; 
this is L$_{\rm{b,max}}$ in the notation of \citet{2010A&A...522A..10C}.}(the 
solid vertical line). \citet{1990A&A...238...95F} determine the LF bump of 
M3 to occur at M$_V$=$-$0.06, \citet{2003PASP..115.1211S} at M$_V$=0.45 
whereas our models reach the composition discontinuity at M$_V$=$-$0.26. 
In the UKRT formula for the diffusion coefficient the change in the surface 
abundances begins almost immediately after the bump. These results differ somewhat from the non-rotating case of  \citet{2010A&A...522A..10C} who find there is a delay of $\sim$ 0.5 magnitudes before the onset of any changes to the surface abundance, 
albeit for different masses and metallicities to those considered here. 
The differences between the two quoted values for the LF bump are due to 
the respective choices of the distance modulus.  We find our models are 
closer to the older results but we refrain from drawing any conclusions. 
A study that includes lower luminosity giants is required in order for us 
to make further comparisons to the observations. 

In Figure \ref{fig:lfbump}b we find that increasing the amount of \el{4}{He} 
in the model has delayed the onset of the bump. These results are consistent 
with \citet{1978IAUS...80..333S} who in addition to this showed that the 
size of the discontinuity is also reduced with increasing \el{4}{He} 
abundance. This is owing to the fact that an increase in \el{4}{He} translates 
to less hydrogen and less of a discontinuity. The fact that the hydrogen 
shell encounters the discontinuity later for models with higher initial 
\el{4}{He} is due to the effect the composition has on the penetration 
of the convective envelope. A higher hydrogen abundance allows for deeper 
penetration of the envelope and hence the shell reaches this depth sooner.      

\section{Conclusion}

M3 is a well studied system that demonstrates the abundance patterns 
we commonly associate with globular clusters. 
Along with many other clusters it displays significant [C/Fe] depletion 
along the RGB, the implication being that some form of internal, non-canonical 
mixing must be occurring. Our models with thermohaline mixing show that 
the carbon and nitrogen observations can be explained if we adopt the 
hybrid theory outlined in \citet{2002PASP..114.1097S}, where stars in 
the cluster are undergoing extra mixing as they ascend the RGB and the 
presence of primordial abundance inhomogeneities due to ON cycling are 
needed to explain the initial carbon and nitrogen abundances. We have used 
observations of M3 to investigate our theoretical understanding of
 thermohaline mixing. Our findings are summarised below: 
   
 \begin{enumerate}

\item The variation of {$\rm{[C/Fe]}$} with magnitude provides a much
      more stringent test of any proposed extra-mixing mechanism
      than simply matching the final \el{12}{C}/\el{13}{C} ratio. When data of
      sufficient quality is available this can constrain the
      details of any proposed extra-mixing formulation. In the
      present case the UKRT formulation of thermohaline mixing
      is a far better fit than the phenomenological prescription
      given by EDL08, although both fit the constraint provided
      by the carbon isotope ratio.

\item The UKRT prescription of thermohaline mixing with  
      \textit{C$_{t}$}=1000 seems to best fit the data for M3. This
      is consistent with the results of \citet{2010A&A...522A..10C} for higher metallicities and CZ07a for a range of metallicities.   

\item We infer that there is a spread of $\sim$0.3 to 0.4 dex
      in [C/Fe] in the stars in M3 from their birth.
      Without this initial difference in [C/Fe] between
      the two populations, thermohaline mixing cannot
      reproduce the change in [C/Fe] seen on the giant
      branch. That there are two populations is absolutely
      required, because once  \textit{C$_{t}$}  is sufficiently large,
      an increase of this coefficient doesn't produce
      a bigger $\Delta$[C/Fe]. Primordial C and N inhomogeneities have been directly observed
     as abundance differences among main sequence stars in globular clusters
     such as M13 and NGC 6752 \citep{2004AJ....127.1579B, 2005A&A...433..597C},
     which have similar metallicity to M3.

\item Thermohaline\footnote{Note, when we refer
   to ``thermohaline mixing" we are referring to the
   linear theory as proposed by \citet{1972ApJ...172..165U}
  and \citet{1980A&A....91..175K}.} mixing can produce the observed values of the
carbon isotopes seen in M3.

\item To reproduce the entire spread of [C/Fe] values
      seen in the giants of M3 it is essential that
      thermohaline mixing operate in both the CN-strong
      and CN-weak populations identified in (3). In this
      case we can explain the full spread in [C/Fe] seen
      near the tip of the giant branch in M3. A similar exercise was carried out by \citet{2003ApJ...593..509D} for the case
   of M92, modelled with a simple parameterized extra-mixing
   formulation. The data for M92 are from many different
   sources and make it difficult to estimate precisely where
   the extra mixing begins. For this reason we do not
   discuss M92 further in this paper.

\item Thermohaline mixing can produce a significant change in [N/Fe] as a
function of M$_V$ on the RGB for initially CN-weak stars but not for initially
CN-strong stars, which have so much N to begin with that
any extra mixing does not significantly affect the surface composition. 

\item The level of depletion of carbon is dependent on the depth to 
which the material is mixed and how fast it is mixed. 

\item Mass loss has little effect on the surface abundances.

\item Both the predicted and observed composition changes take place 
at a luminosity that is higher than the LF bump in \citet{1990A&A...238...95F}. 
The observed abundances begin to decrease at a luminosity lower than 
the LF bump preferred by \citet{2003PASP..115.1211S}. Uncertainties 
in the distance modulus make it difficult to draw further conclusions.  
\end{enumerate}

We have seen that the linear theory of \citet{1972ApJ...172..165U}
  and \citet{1980A&A....91..175K} provides a fit to the
  carbon and nitrogen abundances in the giants in M3 (assuming there
  are two  different populations initially). CZ07a
  have shown that the same theory (and indeed the
  same parameter \textit{C$_{t}$}=1000) seems to fit field stars
  of a range of metallicities (see also \citealt{2010A&A...522A..10C}). It is important to note that
  thermohaline mixing is more than a theory with an
  adjustable parameter. For example, its beginning is
  determined clearly by the fusion of \el{3}{He} which produces
  a molecular weight inversion. Also, the hydrodynamics
  provides the physical formulation for the diffusion
  co-efficient used, and hence its variation throughout
  the star, to within a constant which depends
  on the geometry of the fingers expected in the mixing
  process. Nevertheless, until we have a complete
  theory that also determines this geometric factor,
  or at the least some numerical simulations, there
  is a gap between understanding the fundamental physics
  and the sort of work presented here, to match the
  observations. To this end we note the work of \citet{2010arXiv1006.5481D} which addresses this issue. There has been recently considerable progress in modelling the oceanic case \citep{2010arXiv1008.1808S,2010arXiv1008.1807T}. We are ourselves working on 3D hydrodynamic models in the stellar context, which will be the subject of another paper.

Although our models can explain M3 very well the question remains 
whether this work can be extended to all clusters. The UKRT prescription 
for thermohaline mixing appears to model the internal mixing of a young 
metal rich cluster. Old metal-poor clusters such as M92 will have undergone 
a different mixing history. Furthermore \citet{1978IAUS...80..333S} suggests 
changing the metallicity and helium content will drastically alter the 
location and size of the LF bump. Preliminary work by \citet{2010arXiv1006.5828A} will 
be the subject of subsequent studies. This work further supports the ability 
of thermohaline mixing to explain extra mixing on the RGB.   

\section{Acknowledgements}
We would like to thank Lionel Siess for helping implement the UKRT mixing 
by providing us with his matrix solver. We
also thank Alessandro Chieffi for providing us with his luminosity to
magnitude conversion routine.
We extend our gratitude to Christopher 
Tout for his engaging and helpful discussions on the difference between the 
two mixing schemes. GCA wishes to thank Matteo Cantiello for his time and subsequent discussions on thermohaline mixing. He also wishes to acknowledge the financial support of the APA scholarship. RPC wishes to acknowledge The Wener 
Gren Foundation for their financial support. RJS is funded by the Australian 
Research Council's Discovery Projects Scheme under grant DP0879472. This work was 
supported by the NCI National Facility at the ANU.

\bibliographystyle{apj}

\label{lastpage}

\end{document}